\documentclass[12pt]{article}
\usepackage{latexsym}
\usepackage{graphicx}

\catcode`\@=11
\@addtoreset{equation}{section}

\catcode`@=12
\relax

\newcommand{\nn}{\nonumber}

\begin{document}
\thispagestyle{empty}

\renewcommand{\thefootnote}{\fnsymbol{footnote}}
\font\csc=cmcsc10 scaled\magstep1
{\baselineskip=14pt
 \rightline{
 \vbox{\hbox{TIT-HEP-490}
       \hbox{January 2003}
       \hbox{hep-th/0301226}
}}}

\vfill
\begin{center}
{\LARGE\bf
 Effective Superpotentials for SO/Sp}
\\
\vspace{3mm}
{\LARGE\bf with Flavor 
}
\\
\vspace{3mm}
{\LARGE\bf from Matrix Models
}
\vspace{5mm}
\\

\vfill

\textsc{ \large  Yutaka Ookouchi}\footnote{
      e-mail address : ookouchi@th.phys.titech.ac.jp} and 
\textsc{ \large Yoshiyuki Watabiki}\footnote{
      e-mail address : watabiki@th.phys.titech.ac.jp}\\
\vskip.1in

\textit{\large \baselineskip=15pt
\vskip.1in
  Department of Physics, 
  Tokyo Institute of Technology,\\
  Tokyo 152-8511, Japan
}

\end{center}
\vfill

\begin{abstract}
{\normalsize 
We study matrix models related to $SO/Sp$ gauge theories with flavors. We give the effective superpotentials for gauge theories with arbitrary tree level superpotential up to first instanton level. For quartic tree level superpotential we obtained exact one-cut solution. We also derive Seiberg-Witten curve for these gauge theories from matrix model argument.}

\end{abstract}




\vfill

\setcounter{footnote}{0}
\renewcommand{\thefootnote}{\arabic{footnote}}
\newpage
\setcounter{page}{1}


\section{Introduction}

Recently Dijkgraaf-Vafa proposed that a holomorphic information in 
$\mathcal{N}=1$ gauge 
theories with classical gauge groups derive from matrix model \cite{DV1,DV2,DV3,DV4}. 
In particular the effective superpotentials
for the gauge theories
are given by the genus zero free energies which are the summations of
planar diagrams in matrix models. This perturbative sum of planar
diagrams turns out to be a nonperturbative sum over fractional
instantons in the ${\cal N}=1$ gauge theory. 

Dijkgraaf and Vafa have reached this duality via string theory route using the model with one adjoint matter. It was discussed that in \cite{Vafa,CIV} large $N$ dual of $U(N)$ 
$\mathcal{N}=2$ gauge theory deformed
 by certain tree level superpotential is realized as type IIB string theory on 
Calabi-Yau 
threefold with fluxes.  The discussion for type IIB superstring was extended in \cite{CIV,CKV,CFIKV,EOT,FO,DOT1,DOT2,DOPT}. The equivalence of effective
superpotential for ${\cal N}=1$ gauge theory with that of type IIB
superstring theory on Calabi-Yau manifold with fluxes \cite{Gukov,GVW,TV} was proved in \cite{CV,Ookouchi}.

There are many discussions on this D-V conjecture [19]-[73] .
The generalization of D-V duality to the 
gauge theories 
with massive flavor was discussed in \cite{ACFH,McGreevy,Suzuki1,BR,Feng1,Feng2,Gorsky,DJ1,DJ2,Tachikawa1,Tachikawa2,CDSW,Seiberg}. In \cite{NSW1,NSW2} perturbative analysis of the matrix model, which corresponds to U(N) gauge theories with arbitrary tree level superpotentials. $\mathcal{N}=2$ information was derived these perturbative results including the Seiberg-Witten curve. The analysis beyond the planar limit was discussed in \cite{KMT,DST}. In this note we apply these perturbative result with $SO/Sp$ gauge theories.

In \cite{INO} $SO/Sp$ gauge theories were studied using the superspace method that was developed in \cite{DGLVZ}. In \cite{ACHKR,JO} antisymmetric matrix model was considered as a generalization of DV to $SO/Sp$ gauge theory. On the other hand real symmetric matrix model was used in \cite{FO2,Feng3}. 

The organization of this paper is as follows: In section 2, we review the discussion in \cite{ACHKR} and generalize to the case with massive flavor. In section 3 we give perturbative result for the effective superpotential of $SO/Sp$ gauge theory in terms of the relations discussed in section 2. In section 4 we give exact solution for the case with quartic tree level superpotential with massive matter. In section 5 from the matrix context we derive Seiberg-Witten curves for the gauge theories.

\noindent
Note added:  After completion of this note, we have received \cite{NewAN}, which have some overlaps this paper.

\section{Relations among Free Energies \label{matrix}}
In this section we represent some relations between free energies. These relations are used following section for the computation of effective superpotentials.

\subsection{General Property}
In the standard large $N$ expansion the partition function of matrix model has topological expansion,
\begin{equation}
Z=\mathrm{exp} \left(\sum g_s^{-\chi}{F}_{\chi} \right),
\end{equation}
where $\chi$ is Euler number of the surface. In this letter we will take into account the surface with crosscap, because we discuss $SO/Sp$ gauge theory. Euler number with crosscap $c$ is described as $\chi=2-2g-b-c$, where $g,b,c$ are the number of genus, boundaries and crosscaps respectively. The leading contribution comes from $S^2$ with Euler number $\chi=2$. We denote this free energy as $\mathcal{F}_0$. The Euler number of $RP^2$ is $\chi=1$ that is equal to the Euler number of orientable one boundary surface, disk topology. We denote this one crosscap contribution and one boundary contribution as 
\begin{eqnarray}
F_{\chi=1}=\mathcal{G}_0+\mathcal{F}_1,\quad
\mathcal{G}_0=g_s \log Z|_{RP^2},\quad \quad \mathcal{F}_1=g_s \log Z|_{\mathrm{disk}}.
\end{eqnarray}

Now let us define matrix models corresponding to $SO/Sp$ gauge theories. The partition function of matrix models is given by,
\begin{eqnarray}
Z=\int d\Phi \ \mathrm{exp} \left[-\frac{1}{g_s}W_{0}(\Phi)-\frac{1}{g_s}\sum_{I=1}^{N_f} \left(\tilde{Q}_I\Phi Q^I+m_I\tilde{Q}_IQ^I \right) \right], \label{partition}
\end{eqnarray}
where $\Phi$ is $2M\times 2M$ real antisymmetric matrix\footnote{Note that matrix size $2M$ have not relation to gauge group rank.} or real matrix with relation, $\Phi_{mn}=(J\Phi J)_{nm}$ where $J$ denotes the $Sp$ invariant skew symmetric form
\begin{eqnarray}
\left( 
  \begin{array}{cc}
  \mathbf{0} & \mathbf{1}_{M\times M} \\
  -\mathbf{1}_{M\times M} & \mathbf{0}
  \end{array}
\right), \quad J^2=-\mathbf{1}_{2M\times 2M}.
\end{eqnarray}
$\tilde{Q},Q$ is $2M$ dimensional vector. For simplicity we dropped volume factor for the gauge group, which gives Veneziano-Yankilowicz term. We assume that tree level superpotential $W_0$ is even function, i.e.
\begin{eqnarray}
W_0=\sum_{k=1}^{n+1}\frac{g_{2k}}{2k}\mathrm{Tr}\Phi^{2k}.
\end{eqnarray}

On these two matrix models and Hermitean matrix model that correspond to $U(N)$ gauge theory, some relations are already known. For the free energy coming from disk topology
\begin{eqnarray}
\mathcal{F}_1^{SO}(g)= \mathcal{F}_1^{Sp}(g) =\mathcal{F}_1^{U}(g^{\prime}=2g).
\end{eqnarray}
where gauge groups denote the matrix models corrsponding to these gauge theories. The argument $g,g^{\prime}$ denote coefficients of tree level superpotential. Note that for $SO/Sp$ case the tree level superpotentials are even function then using this relation we assume that the matrix model corresponding to $U(N)$ gauge theory also have even function tree level superpotential. And for genus zero free energy we have
\begin{eqnarray}
 \mathcal{F}_0^{SO}(f)=\mathcal{F}_0^{Sp}(f)=\frac{1}{2}\mathcal{F}_0^{U}(g=2f).
\end{eqnarray}
Multipicating $1/2$ over all factor to $\mathcal{F}_0^U$ and replaceing $W_0$ to $2W_{0}$, namely multipling $2$ to all coefficent of superpotential, it equals to $\mathcal{F}_{0}^{SO}$ and $\mathcal{F}_{0}^{Sp}$. Next relation is the one between $\mathcal{F}_0$ and $\mathcal{G}_0$,
\begin{eqnarray}
\mathcal{G}_0=\mp \frac{1}{2}\frac{\partial \mathcal{F}_0}{\partial S_0},
\end{eqnarray}
where $-$ sign for antisymmetric matrix and $+$ sign for the matrix corresponding to $Sp$ gauge theory respectively and $S_{0}$ is 'tHooft coupling that defined next subsection. These relations were checked in \cite{INO} perturbatively in the viewpoint of field theory using superspace formalism and in \cite{ACHKR,JO} using loop equation for antisymmetric matrix models. We give another derivation for these relations. In \cite{JO} the relation between the $\mathcal{F}_0$ corresponding to the $U(N)$ gauge theory and the one $SO(N)$ gauge theory.

With these relations we can obtain effective superpotentials for $SO/Sp$ gauge theory from matix models. In \cite{INO,ACHKR,JO} the experssion was given,
\begin{eqnarray}
W_{\mathrm{eff}}=\sum_{i=0} N_i\frac{\partial \mathcal{F}_{0}^{SO/Sp}}{\partial S_i}+4 \mathcal{G}_0+\mathcal{F}_1=(N_0\mp 2)\frac{\partial \mathcal{F}_{0}^{SO/Sp}}{\partial S_0}+ \sum_{i=1}N_i\frac{\partial \mathcal{F}_{0}^{SO/Sp}}{\partial S_i}+\mathcal{F}_1.
\end{eqnarray}
It is interesting that two free energy $\mathcal{G}_0$ and $\mathcal{F}_1$ that are the same Euler number, appear to effective superpotential having different coefficient. For the latter convenience let us introduce new notation $\hat{N}_i$, which means that $\hat{N}_0=N_0\mp 2$ and $\hat{N}_i=N_i\ i=1,\cdots$. In section 3 and section 4 we will see effective superpotentials for multicut case and one cut case, using these formulae.

\subsection{Matrix Model for SO(N) with Flavor }

In this subsection we discuss the antisymmetric matrix model. These matrix model  was studied in \cite{ACHKR,JO} for the generalization of DV proposal to the $SO/Sp$ gauge theory. We want to extend these discussion to the model with massive matter. The partition functions of these matrix models are given in (\ref{partition}). As discussed in \cite{McGreevy} we can integrate out $Q$ and diagonalize matrix $\Phi$,
\begin{eqnarray}
\Phi \to UDU^{T}, \qquad D_{i,i+1}=-D_{i+1,i}=i\lambda_i,\ (i=2k-1, k=1,\cdots,N) \ \ \ \mathrm{others\  zero}.\end{eqnarray}
where we used notation $\pm \lambda_i, i=1,\cdots,N$ as eigenvalues of matrix $\Phi$. Then $\lambda_i$ are pure imagenary value. Using these variables $\lambda_i$ we can rewrite as follows,
\begin{eqnarray}
W_{0}(\Phi) \to \sum_{k=1}^{M}2W_{0}(\lambda_k), \qquad
d\Phi \to \left[\prod_{k<l}^M(\lambda_k^2-\lambda_l^2)^2 \right] \prod_i^{M}d\lambda_i.
\end{eqnarray}
\begin{eqnarray}
Z=\int \prod_i^{M}d\lambda_i \mathrm{exp} \left(-\frac{2}{g_s}\sum_{k=1}^MW_{0}(\lambda_k)+2\sum_{k<l} \log (\lambda_k^2-\lambda_l^2)-\sum_{I=1}^{N_f}\sum_i^M \log (\lambda_i^2-m_I^2) \right).
\end{eqnarray}
where third term come from the factor $\mathrm{det}^{-1}(\Phi+m_I)$. The equation of motion for $\lambda_i$ is given by
\begin{eqnarray}
-\frac{2}{g_s}W_0(\lambda_i)+2\sum_{j \ne i} \frac{2\lambda_i}{\lambda_i^2-\lambda_j^2}-\sum_{I=1}^{N_f}\frac{2\lambda_i}{\lambda_i^2-m_I^2}=0. \label{eom}
\end{eqnarray}
As usual let us introduce resolvent $\omega$,
\begin{eqnarray}
\omega \equiv \frac{1}{2M}\mathrm{Tr} \left(\frac{1}{x-\Phi} \right)= \frac{x}{M}\sum_{i=1}^M\frac{1}{x^2-\lambda^2_i}.
\end{eqnarray}
In the large $M$ limit with $g_sM\equiv S$ fixed, it is convenient to introduce density of eigenvalues,
$\rho (\lambda)=\frac{1}{M} \sum_i \delta (i\lambda-i\lambda_i), \int \rho (\lambda)d\lambda =1$, where $i$ is necesary in the delta function becauce $\lambda_i$ are pure imaginary. With this new variables we can write loop equation as follow,
\begin{eqnarray}
\omega (x)^2-\frac{2}{S}\omega (x)W_0^{\prime}(x)+\frac{1}{S^2}f(x)=0,\ \ \ \ f(x)= \frac{2S}{M} \sum_i\frac{\lambda_i W^{\prime}(\lambda_i)-xW^{\prime}(x)}{\lambda^2_i+x^2},
\end{eqnarray}
where third term in (\ref{eom}) supressed in this limit. As in \cite{DV3}, introducing new variable, $y=S \omega(x)+W_0^{\prime}(x)$, we can rewrite loop equation as hyperelliptic curve,
\begin{eqnarray}
y^2={W^{\prime}_0}^2-f(x).
\end{eqnarray}
Generally this Riemann surface has $2n+1$ cuts, we can express the filling number of eigenvalues in terms of the integral around the cut,
\begin{eqnarray}
M_i=M\oint_{\mathrm{i^{th} cut}}y\  dx, \qquad S_i\equiv g_s M_i, \qquad S_0=\frac{g_s}{2}M_0. \label{cutint}
\end{eqnarray}
Since $M_i$'s have $2M=M_0+2\sum_iM_i$, these $S_i$'s have the relation, $S=S_0+\sum_iS_i$.
We can rewrite partition function up to constant term,
\begin{eqnarray}
Z=\mathrm{exp} \left(-\frac{2g_s M}{g_s^2}\int d \lambda \rho(\lambda)W_{0}(\lambda)+\frac{g_s^2M^2}{g_s^2}\int d \lambda d \mu \rho (\lambda) \rho (\mu) \log (\lambda^2-\mu^2) \right. \nn \\ 
\left. -\frac{g_sM}{g_s}\sum_{I=1}^{N_f}\int d \lambda \rho (\lambda) \log (\lambda^2-m_I^2) \right) \label{pertitionfunc}
\end{eqnarray}
Then we can read off the contribution from disk from (\ref{pertitionfunc}) as follows,
\begin{eqnarray}
\mathcal{F}_1=-\frac{S}{2}\sum_{I=1}^{N_f}\int d \lambda \rho (\lambda) \log (\lambda^2-m_I^2).
\end{eqnarray}
We can rewrite this integral as follow,
\begin{eqnarray}
\mathcal{F}_1&=& -\frac{S}{2}\sum_{I=1}^{N_f}\int d \lambda \rho (\lambda)\int_{m_I}^P \frac{2x}{\lambda^2-x^2}-\frac{S}{2}\sum_{I=1}^{N_f}\int d \lambda \rho (\lambda) \log (\lambda^2-P^2) \nonumber \\
&=&-S\sum_{I=1}^{N_f}\int_{-m_I}^P dx \omega (x)-\frac{S}{2}N_f D(P) \\
&=&-\sum_{I=1}^{N_f}\int_{-m_I}^{P}y\ dx +\frac{N_f}{2} W_0(P)-\sum_I^{N_f}\frac{1}{2}W_0(m_I)-\frac{S}{2}N_fD(P) 
\end{eqnarray}
where we defined function $D(P)\equiv \int d\lambda \rho (\lambda)\log (\lambda^2-P^2)$ and $P$ is one of two infinity on the $x$-plane.

\section{Perturbative Analysis}
In this section we give effective superpotential for the $SO/Sp$ gauge theories with arbitrary tree level superpotential up to first instanton correction. In the previous section we study the relation between real symmetric matrix model and unitary matrix model. Using the relation we can obtain real symmetric matrix model result form unitary matrix model result and explicitly give effective superpotential for the $SO/Sp$ gauge theories. Since in \cite{NSW2} perturbative expression for the genus zero free energy and disk amplitude of unitary matrix model was given, we can give the ones for real symmetric matrix model. For this purpose we rewrite equations in \cite{NSW2} for the case with even tree level superpotential and symmetric distribution of eigenvalues.
We write the tree level superpotential as
\begin{eqnarray}
W_0(\Phi)=\sum_{k=1}^{n+1}\frac{g_{2k}}{2k} \mathrm{Tr}\Phi^{2k},\quad W_0^{\prime}(x)=g_{2n+2}x\sum_{k=1}^n(x^2-a_k^2).
\end{eqnarray}
Let us introduce new variables $e_i$,
\begin{eqnarray}
e_0=0 ,\quad e_i=a_i,\quad e_{-i}=-a_i,\quad i=1,2,\cdots n.
\end{eqnarray}
We evaluate matrix integral perturbatively about and extremal point $\Phi=\Phi_0,Q_0=0,\tilde{Q}_0=0$.
\begin{eqnarray}
\Phi_0 = \pmatrix{ e_{-n} \mathbf{1}_{M_n}& 0& \cdots& 0 \cr
                                 0& e_{-n+1} \mathbf{1}_{M_{n-1}}& \cdots& 0 \cr                                 \vdots& \vdots& \ddots& \vdots \cr
                                 0& 0& \cdots&  e_n \mathbf{1}_{M_n} }  ,
\end{eqnarray}
\begin{eqnarray}
S_i\equiv g_s M_i=S_{-i}, \quad g_s\to 0, \quad S_i=\mathrm{fixed}.
\end{eqnarray}
We fix the gauge $\Psi_{ij}=0\ (i\ne j)$and introduce Grassmann-odd ghost matrices $B_{ij}$ and $C_{ij}$. The partition function is given by the gauge-fixed integral
\begin{eqnarray}
Z=\frac{1}{\mathrm{vol}(G)}\mathrm{exp}\left(-\frac{1}{g_s}W_0(\Phi_0) \right)\int d\Psi_{ii}dB_{ij}dC_{ij}dQ^Id\tilde{Q}_I \mathrm{exp}(S_{\mathrm{kin}}+S_{\mathrm{int}}),
\end{eqnarray}
\begin{eqnarray}
S_{\mathrm{kin}}=-\frac{g_{2n+2}}{g_s}\sum_{i=-n}^{i=n}\frac{R_i}{2}\mathrm{Tr}\Psi_{ii}^2-\sum_{i=-n}^n\sum_{j\ne i}e_{ij}\mathrm{Tr}(B_{ji}C_{ij})-\frac{1}{g_s}\sum_{i=-n}^n\sum_{I=1}^{N_f}f_{iI}\tilde{Q}_{iI}Q^I_i,
\end{eqnarray}
\begin{eqnarray}
S_{\mathrm{int}}&=&-\frac{g_{2n+2}}{g_s}\sum_{i=-n}^{n}\sum_{p=3}^{2n+2}\frac{\gamma_{p,i}}{p}\mathrm{Tr}\Psi_{ii}^p \nn \\
&&-\sum_{i=-n}^n\mathrm{Tr}(B_{ji}\Psi_{ii} C_{ij}-B_{ji}C_{ij}\Psi_{jj})-\frac{1}{g_s}\sum_{i=-n}^n\sum_I^{N_f}\tilde{Q}_{iI}\Psi_{ii}Q_i^I ,\label{matrixint}
\end{eqnarray}
where $R_i=\prod_{j\ne i}e_{ij}$ with $e_{ij}=e_i-e_j$, $f_{iI}=e_i+m_I$ and 
\begin{eqnarray}
\gamma_{p,i}=\frac{1}{(p-1)!}\left[\left(\frac{\partial}{\partial x} \right)^{p-1}\prod_{k=-n}^{n}(x-e_k) \right]\bigg|_{x=e_i}.
\end{eqnarray}
Note that these relations are the one for the unitary matrix model. Then the effective superpotential given by this matrix model is the one for $U(N)$ gauge theory. However we can obtain real symmetric matrix model free energy replacing $g_{2n+2}$ to $2g_{2n+2}$ and multiplying the over all factor $1/2$.

\subsection{Genus zero free energy}
Using the relation between $\mathcal{F}_0^{SU}$ and $\mathcal{F}_0^{SO}$ discussed in section 2, we can get the following result,
\begin{eqnarray}
2g_{2n+2}\mathcal{F}_0^{(3)}=\frac{1}{2}\left[(\frac{1}{2}+\frac{1}{6})\sum_{i=-n}^n \frac{S_i^3}{R_i}\left(\sum_{k\ne i}\frac{1}{e_{ik}} \right)^2-\frac{1}{4}\sum_{i=-n}^n \frac{S_i^3}{R_i}\sum_{k\ne i}\sum_{l\ne i,k}\frac{1}{e_{ik}e_{il}}\right. \nn \\ 
\left.-2\sum_{i=-n}^n \sum_{k\ne i}\frac{S_i^2S_k}{R_ie_{ik}}\sum_{l\ne i}\frac{1}{e_{il}}+2\sum_{i=-n}^n \sum_{k\ne i}\sum_{l\ne i}\frac{S_iS_kS_l}{R_ie_{ik}e_{il}}-\sum_{i=-n}^n \sum_{k\ne i}\frac{S_i^2S_k}{R_ie_{ik}^2}\right].
\end{eqnarray}
where we denoted free energy of third order in $S$ as $\mathcal{F}_0^{(3)}$. These result comes from the following Feynman diagrams.


\begin{figure}[htbp]
\begin{center}
\includegraphics[width=12cm,height=3.5cm]
{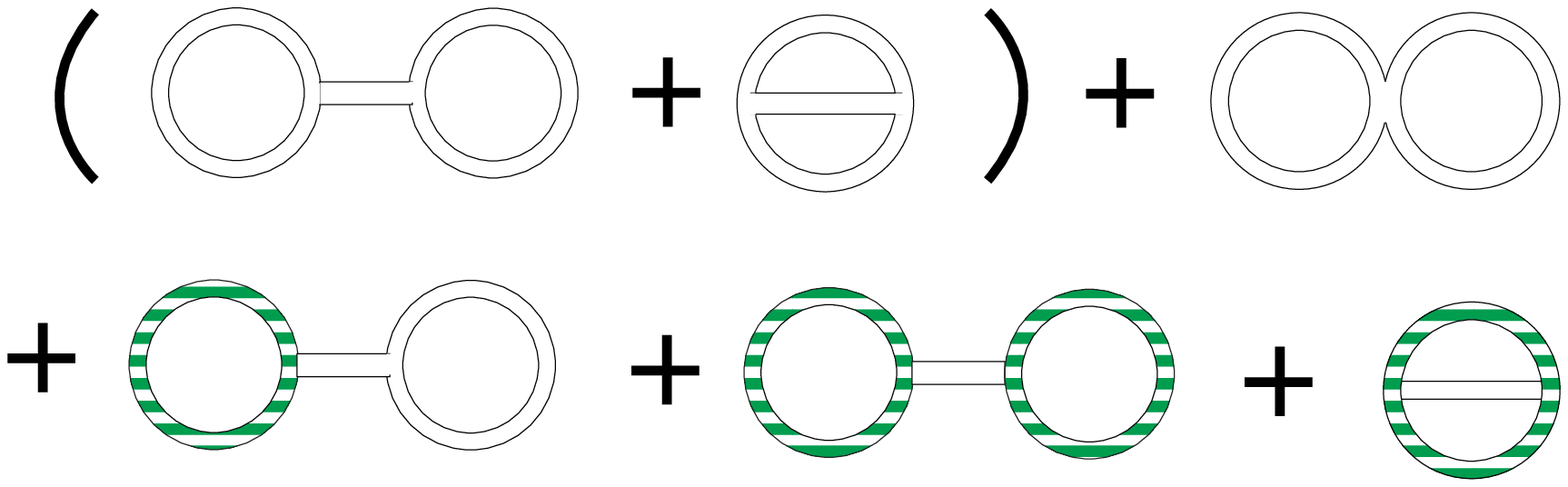}
\caption{\textsf{{\footnotesize 2-loop diagrams contributing to the third order of $S$.  Striped double lines correspond to ghost propagators. Note that we does not compute unoriented diagram but modify the result for Hermitean matrix model with even function tree level superpotential.}}}
 \label{planar}
\end{center}
\end{figure}

Though we are considering the tree level superpotential, which is even functions polynomial, perturbative expansion around classical solution gives potential, which is not even function, (\ref{matrixint}). Thus we have cubic interactions in Feynman diagrams. 

We can check this result for the special case, i.e. $\Phi^4$ case. Let $0,\pm a_1$ are the critical point for the tree level superpotential. Thus $e_{-1}=-a_1,e_-=0,e_1=a_1$ and $R_{-1}=2a_1,R_0=-a_1^2,R_{1}=2a^2$,
\begin{eqnarray}
\mathcal{F}_0^{(3)}&=&\frac{1}{4}\left[\frac{3}{2}\frac{S_1^3}{a_1^4}+\left(-\frac{S_0^3}{2a_1^4}-\frac{S_1^3}{4a_1^4} \right)+\left(-\frac{3S_0S_1^2}{a_1^4}-\frac{3S_1^3}{2a_1^4} \right)\right. \nn \\ 
&+& \left. \left(\frac{S_0^2S_1}{a_1^4}+\frac{S_0S_1^2}{a_1^4}+\frac{S_1^3}{2a_1^4} \right)+\left(\frac{2S_0^2S_1}{a_1^4}-\frac{S_0S_1^2}{a_1^4}-\frac{S_1^3}{4a_1^4} \right) \right]\nn \\
&=&\frac{1}{4g_{4}a_1^4}\left(-\frac{1}{2}S_0^3-2S_0 S_1^2+4S_0^2 S_1\right). \label{SOresult}
\end{eqnarray}
We will see that this result agree with the one in \cite{FO} obtained by Calabi-Yau with fluxes. We review the result in Appendix of \cite{FO}, computation of period of Calabi-Yau manifold with fluxes. The computation was done in semi classical limit $\Lambda \ll 1$. Then the period of compact cycle that interpreted as glueball superfield was small. The period of noncompact cycle was described as the perturbative expansion of the period of compact cycle up to 4-th order. Using the prepotential perturbative part of the result is written as follows,
\begin{eqnarray}
\Pi_0=\frac{\partial \mathcal{F}_0^{CY}}{\partial S_0},\qquad \Pi_1=\frac{\partial \mathcal{F}_0^{CY}}{\partial S_1},
\end{eqnarray}
\begin{eqnarray}
\mathcal{F}_0^{CY}&=&\frac{1}{4ga_1^4}\left(-\frac{1}{2}S_0^3-2S_0S_1^2+4S_0^2S_1\right)+\frac{1}{8g^2a_1^8}\left(\frac{9}{8}S_0^4-14S_0^3S_1+18S_0^2S_1-4S_0S_1^3\right)  \nn \\ &+&\frac{1}{16g^3a_1^{12}}\left(-\frac{9}{2}S_0^5+\frac{233}{3}S_0^4S_1
+\frac{304}{3}S_0^2S_1^3-\frac{524}{3}S_0^3S_1^2-\frac{40}{3}S_0S_1^4\right)+\cdots .
\end{eqnarray}
First order terms agree with (\ref{SOresult}). In \cite{FO} using these perturbative expansion of periods effective superpotential coming from the flux was given. Then integrating out glueball superfield effective superpotential that was written as $\Lambda$ and parameters in tree level superpotential was described. These superpotential can be compared with confining phase superpotential given purely field theory analysis. A few ensamples show agreement these two result. Then the analysis of matrix model describes precisely field theory result.

\subsection{Disk amplitude}
For the disk amplitude there is no difference for two models, then the perturbative result is given in terms of the results in \cite{NSW2},
\begin{eqnarray}
g_{2n+2}\mathcal{F}_1^{(2)}=\sum_{I=1}^{N_f}\left[\sum_{i=-n}^n \frac{S_i^2}{R_if_{iI}}\sum_{j\ne i}\frac{1}{e_{ij}}-2\sum_{I=1}^{N_f}\sum_{j\ne i}\frac{S_iS_j}{R_ie_{ij}f_{iI}}+\frac{1}{2}\sum_{i=-n}^n\frac{S_i^2}{R_if_{iI^2}} \right].
\end{eqnarray}
These result comes from the following Feynman diagrams,
\begin{figure}[htbp]
\begin{center}
\includegraphics[width=12cm,height=1.7cm]
{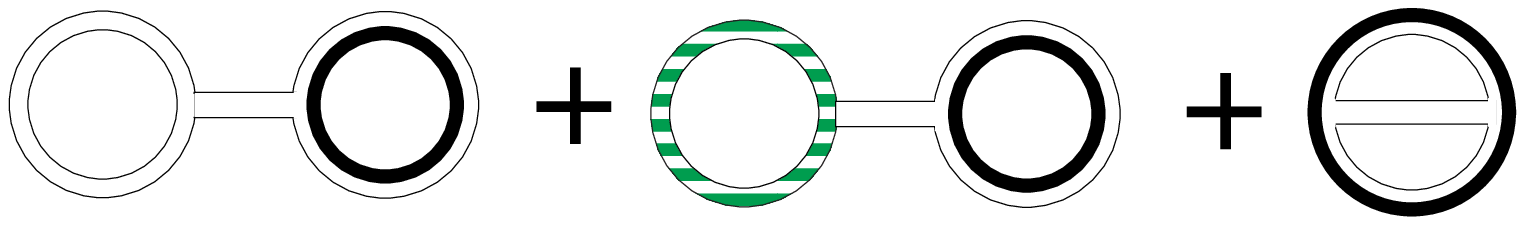}
\caption{\textsf{{\footnotesize 2-loop diagrams with one boundary contributing to the second order of $S$. Fat lines represent a boundary that comes from the propagation of $Q$'s. As in the case $\mathcal{F}_0$ we consider Hermite matrix model and then modify the result for the real symmetric matrix model case.
}}}
 \label{planar2}
\end{center}
\end{figure}

Let us also for this amplitude give explicit representation for the $\Phi^4$ case,
\begin{eqnarray}
g\mathcal{F}_1^{(2)}=\sum_{I=1}^{N_f}\left[-\frac{S_0^2}{2a^2m_I^2}+\frac{2S_0S_1}{a^2(m_I^2-a^2)}+\frac{S_1^2}{(a^2-m_I^2)^2}\right].
\end{eqnarray}

Finally we obtain effective superpotential from matrix model calculation up to second order in $S_i$, namely up to first instanton correction,
\begin{eqnarray}
W_{\mathrm{eff}}=\sum_i^n S_i\left[\log \left( \frac{\Lambda^3}{S_i} \right)^{\hat{N}_i}+1\right]+\sum_{i}^n \hat{N}_i\frac{\partial \mathcal{F}_0^{(3)}}{\partial S_i}+\mathcal{F}_1^{(2)}+\cdots ,
\end{eqnarray}
where first term comes from volume factor \cite{OV}.

\section{Exact Effective Superpotential for Unbroken Gauge Group}
In this section we discuss one-cut solution for $SO/Sp$ case with flavor.  For $\Phi^2$ case there are exact one-cut solutions in \cite{Hofman,Ohta,Ahn}. We want to generalize these results to a quartic tree potential case,
\begin{eqnarray}
 W_{0}=\frac{M}{2}\Phi^2+\frac{g}{4}\Phi^4.  \label{SOgeometry}
\end{eqnarray}
The one-cut Riemann surface is already discussed in \cite{FO},
\begin{eqnarray}
y=\sqrt{{W_{0}^{\prime}}^2+f(x)}=g(x^2+\Delta+2\mu^2)\sqrt{(x-2\mu)(x+2\mu)},
\end{eqnarray}
where $\Delta=\frac{M}{g}$. One of two cut shrinks to be singularity. Using this $y$ we can obtain following relations. It behaves as $1/\lambda$ when $|\lambda|$ goes to infinity. This constraint gives following relation,
\begin{eqnarray}
S=6g\mu^4+2M\mu^2, \qquad \mu^2=-\frac{\Delta}{6}+\frac{\Delta}{6}\sqrt{1+\frac{6S}{g\Delta^2}} \equiv \frac{\Delta}{6}L(S). \label{SOconst2}
\end{eqnarray}
With the relation given in section \ref{matrix} we can calculate the effective superpotential
\begin{eqnarray}
\frac{\partial \mathcal{F}_0}{\partial S}&=&\int_{2\mu}^{\Lambda_0}dx (x^2+\Delta+2\mu^2)\sqrt{x^2-4\mu^2}-W_{0}(P)+\frac{S}{2}D(P) \nn \\
{}&=&-S\log \Lambda_0 +\frac{S}{2}\log \frac{S}{2g}-\frac{S}{4}-\frac{g\Delta^2}{12}L(S)+\frac{S}{2}\log \frac{\Delta}{6}L(S)+\frac{S}{2}D(P),
\end{eqnarray}
\begin{eqnarray}
\mathcal{F}_1&=&\sum_{I=1}^{N_f} \int_{-m_I}^{\Lambda_0}dx (x^2+\Delta+2\mu^2)\sqrt{x^2-4\mu^2}-\sum^{N_f}_I \frac{1}{2}W_{0}(m_I)+\frac{N_f}{2}W(P)-\frac{N_fS}{2}D(P) \nn \\
{}&=&\sum_{I=1}^{N_f} \left[\frac{m_I}{4}(m_I^2-4\mu^2)^{\frac{3}{2}}+\frac{3\mu^2+\Delta}{2}m_I \sqrt{m_I^2-4\mu^2} \right. \nn \\
 &{}&\qquad \quad\left. -2\mu^2(3\mu^2+\Delta) \left(\log 2\Lambda_0-\log (m_I-\sqrt{m_I^2-4\mu^2}) \right)-\frac{1}{2}W(m_I)-\frac{S}{2}D(P) \right] \nn \\
 &=&\sum_{I=1}^{N_f} \left[\frac{m_I^4}{4}K(S)^3+\frac{m_I^2\Delta}{2} \left(1+\frac{L(S)}{2} \right)K(S) \right.  \\
  &{}&\qquad \quad\left. -\frac{\Delta^2}{3}L(S) \left(1+\frac{L(S)}{2} \right) \left(\log \frac{2\Lambda_0}{m_I(1-K(S)) }\right)-\frac{1}{2}W(m_I)-\frac{S}{2}D(P) \right]. \nn
\end{eqnarray}
where $K(S)\equiv \sqrt{1-\frac{2\Delta}{3m_I^2}L(S)}$.

Since $\log \Lambda_0$ divergent piece can be renormalized to bare coupling $\alpha$, we can replace $\Lambda_0$ to $\Lambda$, which is interpreted as dynamically generated energy scale. Until now we used $\Lambda_0$ for simplicity, but correct dimesional parameter is $\Lambda^{\frac{3}{2}}$ so we replace the parameter. And using the matching relation, $\tilde{\Lambda}^{3\hat{N}}=\det m \Lambda^{3\hat{N}-N_f}$, where $m$ is massmatrix that eigenvalues are $m_I$, we can obtain effective superpotential,
\begin{eqnarray}
W_{\mathrm{eff}}&=&S \log \left(\frac{\tilde{\Lambda}^{3\hat{N}}}{S^{\hat{N}}} \right)+\hat{N}\left[-\frac{S}{4}-\frac{g\Delta^2}{12}L(S)+\frac{S}{2}\log \frac{\Delta}{6}L(S) \right] \nn \\
&+&\sum_{I=1}^{N_f} \left[\frac{m_I^4}{4}K(S)^3+\frac{m_I^2\Delta}{2} \left(1+\frac{L(S)}{2} \right)K(S) \right.  \\
  &{}&\qquad \quad\left. -\frac{\Delta^2}{3}L(S) \left(1+\frac{L(S)}{2} \right) \left(\log \frac{2}{(1-K(S)) }\right)-\frac{1}{2}W(m_I) \right].
  \end{eqnarray}

Now let us consider $g \to 0$ limit. There are many discussions for this model with $U(N)$ gauge theory. In this limit we obtain the following result. The effective superpotential for this case is described as 
\begin{eqnarray}
W_{\mathrm{eff}}&=&S \log \left(\frac{\tilde{\Lambda}^{3\hat{N}}}{S^{\hat{N}}} \right)+\hat{N}S  \\
&+&\sum^{N_f}_I \left(-\frac{S}{2}-\frac{Mm_I^2}{4}\sqrt{1-\frac{4S}{Mm_I^2}}+\frac{Mm_I^2}{4}+S\log \left(\frac{1}{2}+\frac{1}{2}\sqrt{1-\frac{4S}{Mm_I^2}} \right) \right). \nn
\end{eqnarray}
This is exactly the same as the result from the discussion for flux with Calabi-Yau geometry in \cite{Ookouchi}. And If we have introduce new function $2f_I(S)\equiv 1+\sqrt{1-\frac{4S}{Mm_I}}$ we can rewrite as,
\begin{eqnarray}
W_{\mathrm{eff}}=S \log \left(\frac{\tilde{\Lambda}^{3\hat{N}}}{S^{\hat{N}}} \right)+\hat{N}S-\sum_I^{N_f}\left(\frac{1}{2}-\frac{1}{2f_I(S)}-\log f_I(S) \right).  \end{eqnarray}
This result agree with matrix model analysis in \cite{Ahn} up to non-ecessial convension of $\Lambda$. 


\section{SW Curves from Matrix Models}
In section 2 we reviewed the knowledge that for introducing the contribution $RP^2$ we have only to replace $N$ to $N\mp 2$. Thus with this lesson we can straight forwardly extend the derivation for Seiberg-Witten curve for the $U(N)$ in \cite{NSW2} to $SO/Sp$ model. To get $\mathcal{N}=2$ information, at first we consider $\mathcal{N}=1$ theory with tree level superpotential is $W_0(x)=\sum_i^{2N+2}\frac{g_{2k}}{2k}\mathrm{Tr}\Phi^{2k}$ and then choose the special vacuume, $N_i=1,\ i=0,1,\cdots ,N$. Next we should turn off the tree level superpotential, namely $g_{2N+2}\to 0$. In this vacuum effective superpotential is given by
\begin{eqnarray}
W_{\mathrm{eff}}&=&(1\mp 2)\frac{\partial \mathcal{F}_0}{\partial S_0}+\sum_{i=1}^N \frac{\partial \mathcal{F}_0}{\partial S_i}-2\pi i \tau_0 \sum_{i=0}^N S_i \nn \\
&=&(1\mp 2)\int_{B_0}y\ dx +\sum_{i=1}^N\int_{B_i} y \ dx -\frac{1}{2}\tau_0 \sum_{i=0}^N \int_{A_i}y\ dx+const  ,
\end{eqnarray}
where $B_i$ is the closed loop through $i$-th and $N$-th cut.

As in $U(N)$ case we use equation of motion for $b_{2i}$'s $i=1,\cdots,N$. The derivative of $y$ with respect to $b_{2i}$ is holomorphic function and form a complete basis. Let us introduce one forms, which form complete basis to holomorphic differentials,\begin{eqnarray}
\frac{\partial y}{\partial b_{2n}}dx =-\frac{1}{2y}x^{2n}dx\equiv \zeta_n .
\end{eqnarray}
Using this notation the equation of motions, $\partial W_{\mathrm{eff}}/\partial b_{2n}=0$, can be written as 
\begin{eqnarray}
\hat{N}\int_{p_0}^{Q}\zeta_k-(\hat{N}-N_f)\int_{p_0}^{P}\zeta_k-\sum_{I=-m_I}^{N_f}\int_{p_0}^{-m_I}\zeta_k=0, \qquad (\mathrm{mod\  period\  integral}).
\end{eqnarray}
Thus from Abel's theorem there exists a function $z(x)$ on the Riemann surface with an $\hat{N}$ th order pole at $Q$, an $(\hat{N}-N_f)$ th zero (or pole if $N_f>\hat{N}$) at $P$, and simple zeros at $-m_I$.
For the $\hat{N}\ge 2N_f$ we find the function $z(x)$ as follow,
\begin{eqnarray}
z(x)&=&A(x) -\sqrt{\left( A(x) \right)^2-C\mathrm{det}_{2N_f}(x+m)},\quad \mathrm{for} \quad SO(2N) \\
z(x)&=&B(x) -\sqrt{B(x)^2-D\mathrm{det}_{2N_f}(x+m)},\quad \ \  \mathrm{for} \quad Sp(2N)
\end{eqnarray}
where $A(x),B(x)$ are airbitrary polynomials defined below and $C,D$ are constant.
These function have $\hat{N}$ th order pole at $Q$ and $\hat{N}-2N_f$ th order zero at $P$ and 1 th order zero at $-m_I$ respectively. For $z$ to be a function on the Riemann surface the square root in $z$ must be proportional to $y$, 
 \begin{eqnarray}
y^2&=& \left[ A(x)^2-C\mathrm{det}_{2N_f}(x+m)\right]{T^{SO}}, \\
y^2&=&\left[B(x)^2-D\mathrm{det}_{2N_f}(x+m)\right]\left({T^{Sp}}\right)^{-1}, \end{eqnarray}
where $T$'s are the proportional function.

We can determine these quantity by taking classical limit $\Lambda \to 0$. In this result matrix model curve $y$ becomes $W_{0}^{\prime}$. So It is natural that $A(x)$ and $B(x)$ are propotional to charactristic function $P_{2N}$, which becomes $W_0^{\prime}$ under the classical limit,
\begin{eqnarray}
A(x)\equiv \frac{P_{2N}}{x^2},\quad B(x)\equiv (x^2P_{2N}(x)+2\Lambda^{4N+4-2N_f} \mathrm{Pf}m).
\end{eqnarray}
For the constants $C,D$ classical value of $Q_I,\tilde{Q}_I$ are zero, it is natural that $C,D$ are proportional to $\Lambda$. With these relation we have the following results in this limit,
\begin{eqnarray}
T^{SO}=x^4, \qquad T^{Sp}=x^2.
\end{eqnarray}
Then we have the following result,
\begin{eqnarray}
y^2= P_{2N}(x)^2-x^4\Lambda^{b_0}\mathrm{det}_{2N_f}(x+m), \\
x^2y^2=\left(x^2P_{2N}(x)+2\Lambda^{2b_0}\mathrm{Pf}m \right)^2-\Lambda^{2b_0}\mathrm{det}_{2N_f}(x+m).
\end{eqnarray}
These are the Seiberg-Witten curve for the gauge theory. Here we have derived these results using only matrix-model methods.

\begin{center}
{\large{\bf Acknowledgements}}
\end{center}
The authors are obliged especially to Hiroyuki Fuji for 
stimulus discussions and useful suggestions.


\end{document}